\documentstyle {article}

\def \R {I \hspace{ - .12truecm} R}
\newcommand { \be} { \begin {equation}}
\newcommand { \en} { \end {equation}}
\newcommand { \bea} { \begin {eqnarray}}
\newcommand { \ena} { \end {eqnarray}} 
\newcommand { \bean} { \begin {eqnarray*}}
\newcommand { \enan} { \end {eqnarray*}}
\newcommand { \bfig} { \begin {figure}}
\newcommand { \efig} { \end {figure}}
\newtheorem {theorem} {Theorem}
\newtheorem {proposition} {Proposition}
\newtheorem {lemma} [proposition] {Lemma}

\begin {document} 

\normalbaselineskip = 5mm \baselineskip = 5mm

\title {Bounded Gain of Energy on the Breathing Circle Billiard 
} 
\author
{
Sylvie Oliffson Kamphorst \thanks { email: syok@mat.ufmg.br }
\ and \
S\^onia Pinto de Carvalho
\thanks{ email: sonia@mat.ufmg.br}
\\
Departamento de Matem\'atica, ICEx, UFMG \\
30123--970 Belo Horizonte, Brasil. 
}

\maketitle 

\begin {abstract}
\noindent
The Breathing Circle is a 2-dimensional generalization of the Fermi Accelerator.
It is shown that the billiard map associated to this model has invariant curves
in phase space, implying that any particle will have bounded gain of energy.
\end {abstract}

\section {Introduction}

The breathing circle is the time-dependent plane region bounded,
instantaneously, by a circle whose radius is a periodic function of time.
The billiard problem on the breathing circle consists on the free motion of a
point particle inside this region, colliding elastically with the moving
boundary.

As J. Koiller, R. Markarian and ourselves have showed in \cite {kn:koi}, time 
varying billiard 
computations can be performed in the same way as done on rigid billiards. But,
beside the  usual coordinates describing static billiards (for the impact point
on the boundary and the direction of the movement), one must also introduce the
energy and time. 
If, for each fixed time, the frozen boundary is a strictly convex plane curve,
the associated time-dependent billiards can be modeled by a 
4-dimensional, volume preserving diffeomorphism which maps successive impacts 
with the moving boundary.

In the special case of the breathing circle, the angular momentum is conserved,
implying a reduction of the model to a 2-dimensional diffeomorphism.
Then, using a corollary of Herman's
{ \sl Th\'eor\`eme des Courbes Translat\'ees } \cite {kn:dou} we prove 
that if the motion of the moving boundary is sufficiently smooth then, for any
admissible initial condition,
the particle will move with bounded velocity and therefore will have a bounded
gain of energy.

A similar case was studied by Levi in (\cite {kn:levi} \S 1), where he shows 
examples of {\sl pulsating soft circular billiards } whose energies stay
bounded for all time.

The study of time-dependent billiards is motivated by two main questions. From 
one side, it is a natural extension to higher dimensions of the Fermi 
Accelerator, studied for instance in \cite {kn:dou}, \cite {kn:ll} and 
\cite {kn:pus}. From another side, among many branches of physics, it is of 
special interest on the study of the motion of particles inside collectively 
excited nuclei. On the attempts to understand the origin of dissipation of 
collective motion in finite Fermi systems (see, for instance, \cite {kn:swia}
and \cite {kn:bur})
models were constructed where the classical part corresponds to time-dependent 
billiards, especially with multipole resonances (with the breathing circle 
corresponding to the monopole mode). 

\section {The breathing circle billiard map}

Let $ R(t) $ be a strictly positive $T$-periodic $ C^k $ function, $k>2$ and 
$ \Gamma (t)=\{ x^2 + y^2 = R^2 (t) \}$ be the (moving) boundary of the 
breathing circle. 
Suppose that at an instant $ t_n $ a particle is on $ \Gamma (t_n) $ at a point
$ (x_n, y_n) $. Let the unitary tangent vector at that point be $ \hat \tau_n $
and the unitary outward normal vector be $ \hat \eta_n $.  
The point on $ \Gamma (t_n) $ is then given by
$ ( x_n, y_n ) = \vec R_n = R_n \, \hat \eta_{n}= R(t_n) \, \hat \eta_{n} $.
Suppose that it leaves the circle from $ (x_n, y_n) $ with velocity
$ \vec v_n $.

It travels on a straight line and hits the boundary again at an instant 
$ t_{n + 1} $ at $ (x_{n + 1} , y_{n + 1} ) = \vec R_{n + 1} $ with
\be
\vec R_{n + 1} = \vec R_{n} + (t_{n + 1} - t_n) \vec v_n 
\label {eq:r}
\en
and will then rebound with velocity 
\be
\vec v_{n + 1} = \vec v_n 
 - 2 < \vec v_n , \hat \eta_{n + 1} > \hat \eta_{n + 1}
 + 2 \left. \frac {d R(t)} {dt}  \right |_{t = t_{n+1} } \hat \eta_{n + 1} \ .
\label {eq:reb}
\en

Let $ \vec L = \vec v \times \vec R $ denote the angular momentum with
respect to the center of the breathing circle, then
$$ 
\vec L_{n + 1} = 
\vec v_{n + 1} \times \vec R_{n + 1} = \vec v_n \times \vec R_{n + 1}
= \vec v_n \times \vec R_{n} = \vec L_n \ .
$$ 
This shows that the angular momentum is conserved at impacts with the moving
boundary.
As it is also obviously conserved between impacts (since the particles
moves on a straight line with constant velocity), angular momentum is
conserved along all the movement.
A first easy consequence of this conservation 
is that a particle never changes orientation with respect 
to the breathing circle, i.e., it always rotates in the same direction.
In other words, if we introduce the notation 
$ L_n = R_n \vec v_n \cdot \hat \tau_n $, is is easy to verify that $L_n$ is 
conserved.

Another consequence of the conservation of angular momentum is that a 
particle never stops if its angular momentum is different from zero.
This is because if $ v_n = | \vec v_n | $, as 
$ L_n = L_0 \neq 0 $ then $ R_n v_n \geq | 
R_n \vec v_n \cdot \hat \tau_n | = |L_0| $ and $ v_n \geq \frac{ |L_0| }{ R_n }
\geq \frac{ |L_0| }{ \overline R } $,
where $ \overline R = $ max $ \{ R(t), t \in [0,T) \} $ .

It also follows that the traveling time of a particle (between two impacts with
the moving boundary) is bounded, because
\be
t_{n+1} - t_{n} = \frac{ \mbox {distance}} { \mbox {velocity}} \leq
\frac {2 \overline R} {v_{n}} \leq \frac {2 \overline R^2} {L_0}
\label {eq:trav}
\en
for every $ n $. 

We introduce the variable $ I_n = -R_n \vec v_n \cdot \hat \eta_n $. It is, in a
certain sense, a natural variable, because it is the normal counterpart of the
conserved quantity $L_n$.

For given $ L_0 \neq 0$ and $I_0 $, $t_0 $ initial impact variables,
let $I_1$ and $t_1$ be the values of $I$ and $t$ at the next impact with the 
breathing circle. Let $ w_n = \frac {d} {dt} R^2 (t) |_{t = t_n}$

\begin {lemma} \label {lem:1}
For any fixed $ L_0 \neq 0 $ and for every $ (I_0, t_0) $ with $ I_0 + 
\frac {w_0} {2} > 0 $, 
the map $ { \cal M } : (I_0, t_0) \mapsto (I_1, t_1) $ is defined by the formulae:
\bea
&& I_1 = I_0 - w_1 + \left( \frac {L_0^2 + I_0^2} {R_0^2} (t_1 - t_0) - 
2 I_0 \right)
\label {eq:I} \\
&& (t_1 - t_0) \left( \frac {L_0^2 + I_0^2} {R_0^2} (t_1 - t_0) - 2 I_0 \right)
= R_1^2 - R_0^2
\label {eq:t}
\ena
\end {lemma}
\begin {proof}
Equation (\ref {eq:t}) above is easily obtained after computing 
$ | \vec R_1 |^2 $  from (\ref {eq:r}). Now, if the particle moves inside the 
breathing circle region, it's normal velocity at the impacts must be bigger 
than the normal velocity of the boundary, i.e., 
$ - \vec v_0 \cdot \hat \eta_0 + R'(t_0) > 0 $ 
or, using the variables introduced above, $ I_0 + \frac {w_0} {2} >0 $.
This condition implies that (\ref {eq:t}) has a solution $ t_1 > t_0 $ and 
thus the map is well defined under this 
assumption. (In \cite {kn:koi} we have in fact proved a little more:
$ I_0 + \frac {w_0} {2} >0 $ implies $ I_1 + \frac {w_1} {2} \geq 0 $.)

On the other hand, from (\ref {eq:r}) we have that 
$ (t_1 - t_0) \vec v_0 \cdot \vec v_0 =( \vec R_1 - \vec R_0) \cdot \vec v_0 $.
This relation together with the rebound formula (\ref {eq:reb}) yields
(\ref {eq:I}).
\end {proof}

Formulae (\ref {eq:I}) and (\ref {eq:t}) above are invariant under the 
translations $ t + nT $, where $ T $ is the period of $ R(t) $ and $ n $ is 
any integer. So, we can take
$ t\, (\hbox{mod } T) \in S^1$ and then the domain of $ { \cal M } $ is contained in 
the cylinder $ \R \times S^1 $.
It is worthwhile to point out that, although time-dependent billiards
are described by four variables, the breathing circle only takes two,
namely $I$ and $ t\, (\hbox{mod } T)$, the other two independent variables
being the (conserved) angular momentum and its conjugate variable.
Moreover, as $ L_0 $ appears only squared in the calculations,
we can suppose $ L_0 > 0 $. The case $L_0=0$ corresponds to the movement on the
diameter of the
breathing circle and is equivalent to the 1-dimensional Fermi accelerator, 
studied, for instance, in \cite {kn:dou}, \cite {kn:ll} and \cite {kn:pus}. 

Now let 
$ (I_1, t_1) = { \cal M } (I_0, t_0) $ and define $ (I_{ - 1},t_{ - 1}) $
by 
$ { \cal M } (I_{ - 1},t_{ - 1}) = (I_0, t_0) $. Let
\bean
A_1 & = & \{ \, (I_0, t_0) \ | \
I_0 + \frac {w_0} {2} > 0 \ { \mbox { and } }  I_1 +
 \frac {w_1} {2} = 0 \, \} \\ 
A_{-1}
& = & \{ \,  (I_0, t_0) \ | \ 
I_0 + \frac {w_0} {2} > 0 \ { \mbox { and } } 
I_{ - 1} + \frac {w_{ - 1}}{2} = 0 \,  \}.
\enan
(With our assumptions on $ R(t) $, the Lebesgue measure on ${\R}^2 $ of 
these sets is zero.)

Let $ D = \R \times S^1 \setminus (A_1 \cup A_{ - 1}) $.

\begin {lemma} \label {lem:2}
If $ R(t) $ is a strictly positive $ T $-periodic $ C^k $ function,
$ k>2 $ then for each fixed $ L_0 \neq 0 $ the map $ { \cal M } : D \to D $,
given by (\ref {eq:I}) and (\ref {eq:t}), is a $ C^{k - 2} $ diffeomorphism, 
preserving the area $ \displaystyle {d\mu = \frac {2 I + w } {2 R^2} dI dt }$. 
\end {lemma}
\begin {proof}
the proof follows from Theorem 1 in \cite {kn:koi}
\end {proof}

\section {Changing coordinates: Some properties of the billiard map}

From the section above, it seems natural to introduce the variable
$ J = 2 I + w $. The advantages of this choice will become clear in
what follows.
In the new coordinates $ (J, t) $ lemma \ref {lem:1} rewrites:

{\bf Lemma \ref {lem:1}'} { \em For a fixed
$ L_0 \neq 0 $ and given an initial condition $ (J_0, t_0) $ with $ J_0 > 0 $,
its iterated by the billiard map, $ (J_1, t_1) $, is given by the formulae:}
\bea
& & J_1 = J_0 - (w_0 + w_1) + 2 \frac {R_1^2 - R_0^2} {t_1 - t_0} 
\label {eq:J} \\
& & 
(t_1 - t_0) { 4 L_0^2 + ( J_0 - w_0 )^2 \over 4 R_0^2 }
=  J_0 - w_0 + { R_1^2 - R_0^2 \over t_1 - t_0 }.
\label {eq:t1}
\ena

Since $ R(t) $ is at least $ C^2 $, there exist $ \xi_0 $ and $ \xi_1 $
such that 
\be
 R_1^2 - R_0^2 = \frac {w_0 + w_1} {2} (t_1 - t_0)
 + \frac {w'( \xi_0) } {4} (t_1 - t_0)^2 - \frac {w'( \xi_1) } {4} 
 (t_0 - t_1)^2,
 \label{eq:int}
 \en 
and
$$ J_1 = J_0 + \frac {w'( \xi_0) + w'( \xi_1) } {2} (t_1 - t_0). $$ 

For $ L_0 \neq 0$ and by inequality (\ref {eq:trav}) we have
\be
| J_1 - J_0 | \leq \overline{w}' \frac { 2 \overline{R}^2} { | L_0 |}
\label{eq:adi}
\en
where $ \overline {w}' =\hbox{max} \{ | w'(t) |, t \in [0,T]\}$.

{\bf Remark}: Suppose that the breathing circle moves slowly, i.e.,
$ R^2 (t) = 1 + \epsilon g(t) $, for a small parameter $ \epsilon $ and $g$ a 
$ C^k $, $T$-periodic function.
Then $ w'(t) = \epsilon g''(t)= { \cal O} ( \epsilon) $.
Let $ J_n $ be the value of $ J $ at the $ {\rm n}^{th} $ impact. 
For $ n = { \cal O}( \frac {1}{ \epsilon}) $ 
$$ | J_n - J_0| \leq n \overline {w}' \frac {2 \overline {R}^2} { |L_0|} =
{ \cal O}(1) $$ 
which shows that $ J $ is an adiabatic invariant (see \cite {kn:arno2} for the 
definition of adiabatic invariant).

Let $ B_ \lambda $ denotes the cylinder $ ( \lambda , + \infty) \times S^1 =
 \{ (J, t), J > \lambda, t \in S^1 \} $.
\begin {lemma} \label {lem:3}
Suppose that $ R(t) $ is a strictly positive $ T $-periodic $ C^k $ function, 
$k>2$. Let
$$ M : B_{ \overline {w}'( \frac {2 \overline {R}^2} { |L_0|} + 1)} 
\longrightarrow M ( B_{ \overline {w}'( \frac {2 \overline {R}^2} { |L_0|} + 
1)}) \subset B_{ \overline {w}'} $$ 
be defined by formulae (\ref {eq:J}) and (\ref {eq:t1}). Then $M$ is a 
$ C^{k - 2} $ diffeomorphism, preserving the measure \break
$ d\mu = \frac {J} {4R^2} 
dJ dt $. 
\end {lemma}
\begin {proof}
For a fixed $ L_0 \neq 0 $, we have automatically from (\ref {eq:adi}) that
$$ J_0> \overline {w}'( \frac {2 \overline {R}^2} { |L_0|} + 1) 
\Rightarrow 
 J_1> \overline {w}'>0 \ .
$$
The proof follows immediately from this fact and lemma~\ref {lem:2}.
\end {proof}

\begin {lemma}
Given any curve $ \Gamma $ in the cylinder 
$ B_{ \overline {w}'( \frac {2 \overline {R}^2} { |L_0|} + 1)}$
homotopic to $ S^1 \times \{0 \} $ then
$ M ( \Gamma) \cap \Gamma \neq \emptyset $. In other words,
$ M $ has the property of intersection.
\end {lemma}

\begin {proof}
$ M $ preserves a measure, absolutely continuos with respect to the Lebesgue 
measure. This implies the property of intersection \cite {kn:dou}.
\end {proof}

\section {Invariant Curves}

The purpose of this section is to investigate the behaviour of the
breathing circle billiard map in the neighbourhood of $ \infty $. We shall
show that a map can be defined which exhibits some properties of twist
maps. More specifically, using a result 
proved by R. Douady \cite {kn:dou} we will prove the existence of
invariant curves for values of $J$ sufficiently large.

For a fixed $ L_0 \neq 0 $, and since
$$ J_0> \overline {w}'( \frac {2 \overline {R}^2} { |L_0|} + 1) 
\Rightarrow 
 J_1> \overline {w}'>0 \ 
$$
the change of variables 
$ (z, t) = ( \frac {1} {J}, t) $ is well defined on 
$ B_{ \overline {w}'( \frac {2 \overline {R}^2} { |L_0|} + 1)} $ and is 
$ C^ \infty $ .

Let 
$0 < \delta \leq \frac { |L_0| } { \overline {w}'(2 \overline {R}^2 + |L_0|)} $.
Then the map 
$$
\begin {array} {c c c c}
N: & (0, \delta) \times S^1 & \to & {I \hspace{ - .12truecm} R}^ + \times S^1\\
   & (z_0, t_0) & \mapsto  &  M( 1/z_0, t_0) \\
\end {array}
$$
defined by
\bea
z_1 = z_0 \frac {1} {1 + g(t_1, t_0) z_0} \, \, \mbox {with} \, \,
g(t_1, t_0) 
= - (w_0 + w_1) + 2 \frac {R_1^2 - R_0^2} {t_1 - t_0} \label {eq:z} \\
\left[ 4 L_0^2 z_0^2 + (1 - w_0 z_0)^2 \right] 
(t_1 - t_0) + 4R_0^2 z_0 (w_0 z_0 - 1)
 = 4 R_0^2 z_0^2 \frac {R_1^2 - R_0^2} {t_1 - t_0} \label {eq:t2}
 \ena
is a $ C^{k - 2} $ diffeomorphism, with the property of intersection.

\begin {lemma} \label {lem: N}
$ N $ can be extended to a $ C^{k - 2} $ diffeomorphism $ \tilde N $ 
on a neighbourhood of $ \{0 \} \times S^1 $. $ \tilde N $ has the property of 
intersection
and a development of the form:
\bea
&z_1 = & z_0 + { \cal O} (z_0^2) \nonumber \\
&t_1 = & t_0 + 4 R_0^2 z_0 + { \cal O} (z_0^2) \nonumber
\ena
\end {lemma}
\begin {proof}
First of all, let us show that $ t_1 = t_1 (t_0, z_0) $ can be extended
for $ z_0 < 0 $. $ R(t) $ being $ C^k $, $k>2$ then
$ f( t_1 - t_0)= R_1^2 - R_0^2 - w_0 (t_1 - t_0 ) $ is a
$ C^{k - 1} $ function with $ f(0) = f'(0) = 0 $.
By Hadamard's Lemma
$ f( t_1 - t_0) = (t_1 - t_0) h(t_1 - t_0) $, where
$ h $ is a $ C^{k - 2} $ function.
Using (\ref {eq:t2}) and discarding the trivial solution 
$ t_1 \equiv t_0 $, we have:
$$ \left[ 4 L_0^2 z_0^2 + (1 - w_0 z_0)^2 \right] (t_1 - t_0) +
4R_0^2 z_0 (w_0 z_0 - 1) = 4 R_0^2 z_0^2 \left[ w_0 + h(t_1 - t_0) \right]. $$ 

Let
$$ F(t_0, t_1, z_0) = \left[ 4 L_0^2 z_0^2 + (1 - w_0 z_0)^2 \right]
 (t_1 - t_0) + 4R_0^2 z_0 (w_0 z_0 - 1) - 4 R_0^2 z_0^2 
\left[ w_0 + h(t_1 - t_0) \right]. $$ 
$ F $ is a $ C^{k - 2} $ function on $ S^1 \times S^1 \times 
{I \hspace{ - .12truecm} R} $ such that 
$ F(t_0, t_0, 0) = 0 $ and $ \frac{ \partial F}{ \partial t_1} 
(t_0, t_0, 0) = 1 $. If $ k>2 $ and for each fixed $ t_0 \in S^1 $,
by the Implicit Function Theorem we have $ t_1 = t_1 (z_0, t_0) $ on a 
neighbourhood of $ (0,t_0) $.
Moreover, as $ S^1 $ is compact and $ t_1 (0, t_0) = t_0 $ for every $ t_0 $,
we can find a neighbourhood of $ \{0 \} \times S^1 $ where
$ t_1 = t_1 (z_0, t_0) $ is a $ C^{k - 2} $ function.

This implies that $ g(z_0, t_0) = - (w(t_0) + w(t_1(z_0, t_0)))
+ 2 \frac {R^2 (t_1(z_0, t_0)) - R^2 (t_0) } { t_1(z_0, t_0) - t_0} $
has a $ C^{k - 2} $ extension on a neighbourhood of $ \{0 \} \times S^1 $. 
Clearly it is bounded and, perhaps on a smaller neighbourhood, we can take
$ | g(t_1, t_0) z_0|<1 $. 
So, $ z_1(z_0, t_0) $ has a $ C^{k - 2} $ extension on a neighbourhood of 
$ \{0 \} \times S^1 $. 

Analogously, $ N^{ - 1} $ can be extended on a neighbourhood of 
$ \{0 \} \times S^1 $. 
$ N $ is then extended to a $ C^{k - 2} $ diffeomorphism $ \tilde N $ on a 
neighbourhood of $ \{0 \} \times S^1 $. 

Since the formulae remain unchanged, the property of intersection is 
preserved for curves on $ z_0 > 0 $ or $ z_0 < 0 $.
But $ \tilde N(0, t_0) = (0,t_0) $ for any $ t_0 \in S^1 $. 
Then, if a curve $ \Gamma $ cuts the circle $ \{0 \} \times S^1 $,
$ \tilde N( \Gamma) \cap \Gamma \neq \emptyset $, and $ \tilde N $ has 
the property of intersection.

Finally, a simple calculation leads to the development of $ \tilde N $. 
\end {proof}

\begin {proposition}
If
$ k>7 $,
we have a family of invariant curves for $ \tilde N $ that {\sl protect}
$ \{ 0 \} \times S^1 $, in the following sense: Every orbit of $ \tilde N $
with no intersection with $ \{ 0 \} \times S^1 $, stay at a bounded distance,
above and bellow,
from $ \{ 0 \} \times S^1 $. Those bounds are arbitrarily small strictly 
positive constants.
\end {proposition}
\begin {proof}
Lemma \ref {lem: N} give the conditions to apply a corollary of Herman's
{\sl Th\'eor\`eme des Courbes Translat\'ees},
proved by R. Douady ( \cite {kn:dou}, page III - 8) and the result follows.
\end{proof}

\section {Boundedness of the velocity}

Coming back to the coordinates $ (J, t) $ and the diffeomorphism $ M $,
let us denote by $ (J_n, t_n) $ the points of the orbit of 
$ (J_0, t_0) $ under $ M $. 
An admissible initial condition for $ M $ is the set of $ (J_0, t_0) $
such that $ J_n > 0, \forall n $ (which is the condition of existence of the 
next impact given in lemma \ref {lem:1}').

\begin {theorem}
Given the billiard on the breathing circle $ x^2 + y^2 = R(t)^2 $,
with $ R(t) $ a strictly positive $ T $-periodic 
$ C^k $ function, $ k>7 $, then, for any admissible initial condition,
a particle will move with bounded velocity.
\end {theorem}
\begin {proof}
For $ L_0 \neq 0 $, since $ \tilde N $ has a family of invariant curves 
that {\sl protect} $ \{ 0 \} \times S^1 $ and $ \tilde N = N $ for $ z_0 > 0 $,
we have a family of invariant curves of $ M $ that {\sl protect} $ \infty $, 
i.e., there exist a family of functions $ \Gamma_n :
S^1 \to ( \overline {w}'( \frac {2 \overline {R}^2} { |L_0|} + 1), + \infty) $,
approaching infinity with norm $ C^{k - 2} $, whose graphs are invariant under
$ M $. 

Given $ (J_0, t_0) $ such that 
$ \Gamma_l (t_0) < J_0 < \Gamma_m (t_0) $ for some $l$ and $m$,
we will have that
$ 0 < \mbox {inf}_{t \in S^1} \Gamma_l (t) < J_n < 
\mbox {sup}_{t \in S^1} \Gamma_m (t) $,
for every $ n $. Therefore $ (J_0, t_0) $ is an admissible initial condition 
and $ J_n $ will remain bounded.

On the other hand,
if $ (J_0, t_0) $ is bellow all the invariant curves, then $ J_n $ will
be bounded above although it may be equal to zero. 
Anyway, for every admissible initial condition, $ J_n $ will remain bounded.

Since $ J_n = 2 (I_n + \frac {w_n} {2}) =
2( - R_n \vec v_n \cdot \hat \eta_n + \frac {w_n} {2}) $,
it is obvious that $ J_n $ is bounded if and only 
if $ v_n = | \vec v_n | $ is bounded.

If $ L_0 = 0 $, we have the motion along the diameter of the breathing 
circle, that corresponds to the 1-dimensional Fermi accelerator,
which has bounded velocities for any admissible initial condition 
(see, for instance, \cite {kn:dou}).
\end {proof}

{\bf Remark}: This result is also true if we consider the billiard in any higher
dimension breathing sphere $S^n$, instead of the circle $ S^1 $, 
since the conservation of the angular momentum implies that the movement will
occur on a plane and so, the
effective boundary will be the breathing circle.

In terms of the dynamics of the breathing circle billiard, we can translate
Theorem 1 in the following way: For each fixed $ L_0 $, there exists a bound
$ K $, depending on $ R(t) $ and decreasing with $ L_0 $,
such that for $ J> K $,
there are infinitely many spanning curves that may not exist
for $ J<K $. This fact is illustrated on the pictures bellow where we plot the
phase-space for  
$ R(t)=1+\frac{0.1}{4 \pi ^2}\cos 2 \pi t$ and different choices of $L_0$.
(Because of the special form of $ R(t) $,  $ J $ is, in fact, an eternal 
adiabatic invariant and, for big $ J $'s, all curves are invariant.)

\begin {figure} [h]
\vskip  5.5truecm \includegraphics{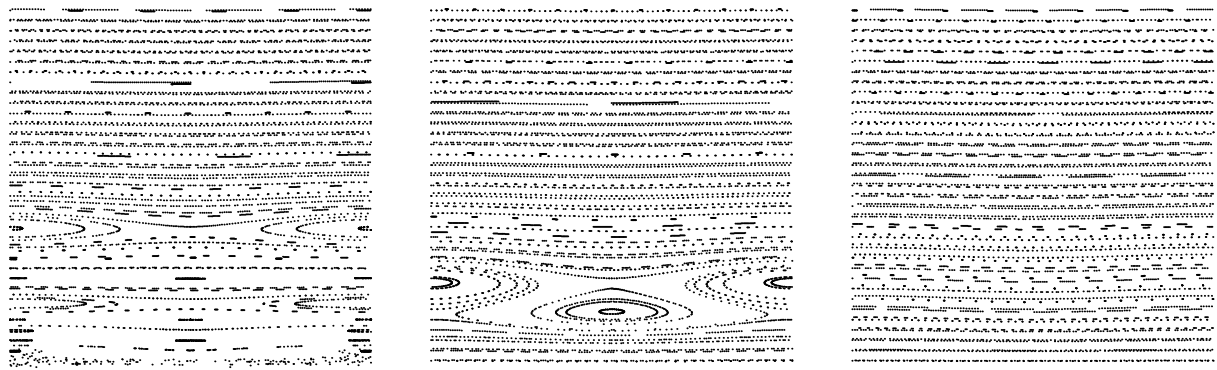}
\caption {Phase-space of the breathing circle for
$t \in [0,1)$, $ J \in [0,8] $, 
and $ L_0 = 0.25, 0.98$ and $1.3$.}
\label{fig:mov}
\end {figure}

Clearly our approach does not apply to the low energy regime (i.e., for small
$L_0$'s and/or small $J$'s). For the 1-dimensional case, a simplified model was
introduced by Lichtemberg and Liebermann \cite {kn:ll}. This approximation
and some comparisons
with the standard map seems to give some valuable insights about this energy
regime. In the simplified
model one allows the boundary to interact with the particle through momentum
exchange (via a given $ w(t) $ periodic in time), but assuming that the boundary
does not change in time and so $ R(t) \equiv R_0 $. In this case, the associated
billiard map is given by
\bean
&& J_1 = J_0 - (w_0+w_1) \\
&& t_1 = t_0  +  \frac { 4 R_0^2} {4 L_0^2 + (J_0-w_0)^2} (J_0-w_0).
\enan
This simplified model allows unphysical negative values of $J$,
which are forbidden in the complete model.

On the next set of pictures we show the phase-space of the simplified model
with the same conditions and $ w(t) $ we took in figure \ref{fig:mov}. Except
for a small neighbourhood of $J=0$, the dynamical behaviour seems astonishingly
similar.

\begin {figure} [h]
\vskip 5.5truecm \includegraphics{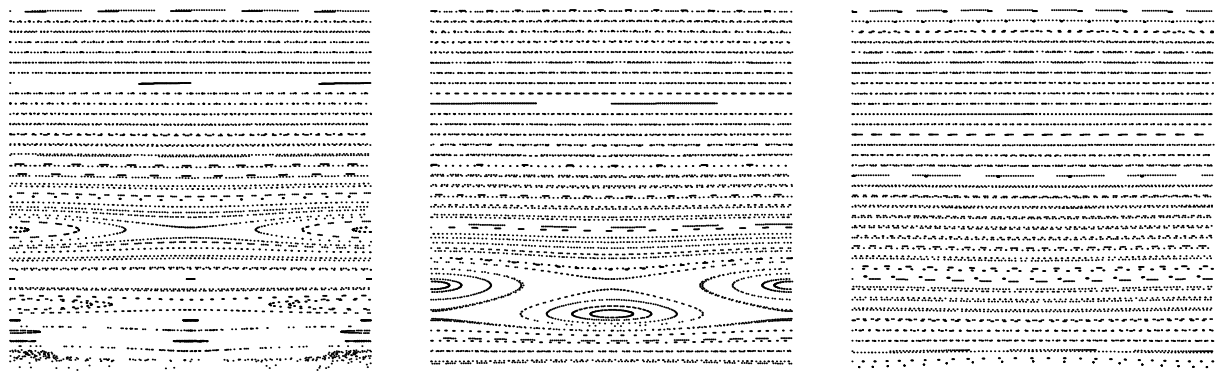}
\caption {Phase-space of the simplified breathing circle for
$t \in [0,1)$, $ J \in [0,8] $, 
and $ L_0 = 0.25, 0.98$ and $1.3$.}
\end {figure}

Finally we point out that the existence of a conserved quantity (the angular
momentum) is the clue of our
proof, since it reduces the dimension to two. Unfortunately, it is easy to prove
that a 2-dimensional moving billiard, with convex frozen boundary and with only
normal deformation, preserves the angular momentum if and only if it is the
breathing circle.
So, the extension of this result to other time dependent billiards will
depend on the existence of some constant of motion or to the extension of the 
techniques to higher dimensions.


\begin {thebibliography} {99}
\bibitem {kn:arno2}
V. I. Arnold: { Small denominators and problems of stability of motion
in classical and celestial mechanics}, Russ. Math. Surveys {\bf 18:6},
 85-191 (1963).
\bibitem {kn:swia}
J. Blocki, Y. Boneh, J. R. Nix, J. Randrup, M. Robel, A. J. Sierk, W. J. 
Swiatecki: One-Body Dissipation and the Super-Viscidity of Nuclei, Annals of 
Physics {\bf 113}-2, 330-386 (1978).
\bibitem{kn:bur}
G. F. Burgio, M. Baldo, A. Rapisarda: Chaoticity in Vibrating Nuclear Billiards,
Phys.Rev.C, {\bf 52}, 2475- (1995).
\bibitem {kn:dou}
R. Douady: Applications du th\'eor\`eme des tores invariants, Th\`ese
de 3\`eme Cycle, Univ. Paris VII (1982).
\bibitem {kn:koi}
J. Koiller, R. Markarian, S. Oliffson Kamphorst, S. Pinto de Carvalho:
Time-dependent billiards, Nonlinearity {\bf 8}, 983-1003 (1995). 
\bibitem{kn:levi}
M. Levi:  Quasiperiodic  motions  in  superquadratic  time-periodic 
potentials, Comm. Math. Phys, {\bf 143}  43-83 (1991).
\bibitem {kn:ll}
A.J. Lichtenberg, M.A. Liebermann: { \em Regular and Stochastic Motion},
Springer-Verlag Appl. Math. Sci. { \bf 38}, Berlin, Heidelberg, New 
York, Springer-Verlag (1983).
\bibitem {kn:pus}
T. Kr\"uger, L.D. Pustyl'nikov, S.E. Troubetzkoy: Acceleration of bouncing balls
in external fields, Nonlinearity, {\bf 8}, 397-410 (1995).\\
L.D. Pustyl'nikov: On Ulam's Problem, Theor. Math. Phys. { \bf 57},
1035-1038 (1983). \\ 
L.D. Pustyl'nikov: On the Fermi-Ulam Model, Soviet Math. Dokl.
{ \bf 35:1}, 88-92 (1987).
\end {thebibliography}
\end {document}